\documentclass[a4paper,fleqn,num]{cas-dc}

\usepackage[numbers,sort&compress]{natbib} 
\usepackage{amssymb}
\usepackage{amsmath}
\usepackage{subfigure}
\usepackage{tabularx}
\usepackage{booktabs}
\usepackage{graphicx}
\usepackage{multirow}
\usepackage{nomencl}
\usepackage{longtable}
\usepackage{array}  

 

\begin{document}

\shorttitle{} 

\shortauthors{}  

\title [mode = title]{Temperature calibration of surface emissivities with an improved thermal image enhancement network}  

\tnotemark[1] 

\tnotetext[1]{This work was supported by the Zhejiang Provincial Special Support Program [Nos. KP202104] and partially by the National Natural Science Foundation of China [No. 62172132].} 

\author[1,2]{Ning Chu}
\ead{chuning1983@sina.com}

\author[3]{Siya Zheng}
\ead{zsy_8020@163.com}

\author[3]{Shanqing Zhang}
\ead{sqzhang@hdu.edu.cn}

\author[3,4]{Li Li}
\ead{lili2008@hdu.edu.cn}

\author[1,2]{Caifang Cai}
\ead{caifang_cai@163.com}

\author[1,2]{Ali Mohammad-Djafari}
\ead{djafari@ieee.org}

\author[1]{Feng Zhao}
\ead{13840424515@163.com}

\author[3]{Yuanbo Song}
\ead{syb20001005@outlook.com}

\affiliation[1]{organization={Ningbo Institute of Digital Twin(IDT)},
                city={Ningbo},
                postcode={315211}, 
                country={China}}

\affiliation[2]{organization={Zhejiang Shangfeng Special Blower Company Ltd.},
                city={Shaoxing},
                postcode={312352}, 
                country={China}}

\affiliation[3]{organization={College of Computer Science and Technology, Hangzhou Dianzi University},
                city={Hangzhou},
                postcode={310018}, 
                country={China}}
                
\affiliation[4]{organization={Hangzhou Dianzi University Shangyu Institute of Science and Engineering Co., Ltd.},
                city={Shaoxing},
                postcode={312399}, 
                country={China}}

\cortext[cor1]{Corresponding author: Shanqing Zhang}

\begin{abstract}
Infrared thermography faces persistent challenges in temperature accuracy due to material emissivity variations, where existing methods often neglect the joint optimization of radiometric calibration and image degradation. This study introduces a physically guided neural framework that unifies temperature correction and image enhancement through a symmetric skip-CNN architecture and an emissivity-aware attention module. The pre-processing stage segments the ROIs of the image and and initially corrected the firing rate. A novel dual-constrained loss function strengthens the statistical consistency between the target and reference regions through mean-variance alignment and histogram matching based on Kullback-Leibler dispersion. The method works by dynamically fusing thermal radiation features and spatial context, and the model suppresses emissivity artifacts while recovering structural details. After validating the industrial blower system under different conditions, the improved network realizes the dynamic fusion of thermal radiation characteristics and spatial background, with accurate calibration results in various industrial conditions.
\end{abstract}





\maketitle

\section{Introduction}

Infrared thermal imaging has revolutionized industrial diagnostics by converting emitted thermal radiation into pseudo-color images and quantitative temperature profiles, overcoming human vision limitations in extreme operational environments \cite{cite0, cite1}. The proliferation of cost-effective thermal cameras \cite{cite1} has enabled widespread adoption for non-contact monitoring of critical equipment under high temperatures, pressures, and dynamic conditions \cite{cite2}. Recent advances in deep learning further enhance this capability, as demonstrated by Resendiz-Ochoa et al. \cite{cite32}, who developed a deep feature learning framework for multi-fault detection in electromechanical systems using thermal imaging. Such innovations underscore infrared thermography's role in predictive maintenance and safety assurance.

\begin{figure}[!t]
\centering
\subfigure[]{
\includegraphics[width=1in]{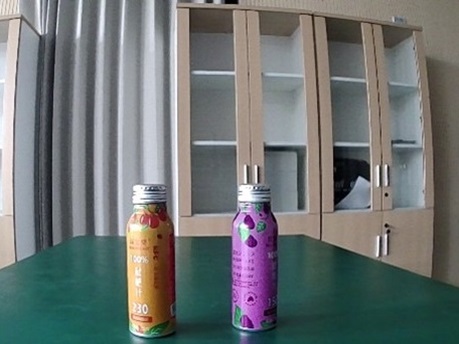}
}
\subfigure[]{
\includegraphics[width=1in]{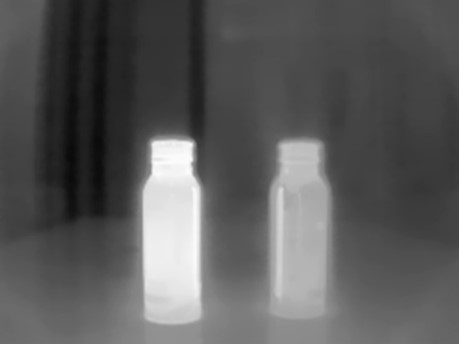}
}

\subfigure[]{
\includegraphics[width=1in]{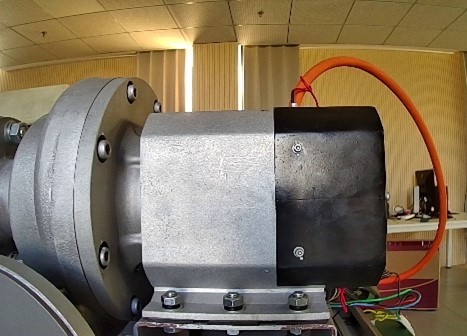}
}
\subfigure[]{
\includegraphics[width=1in]{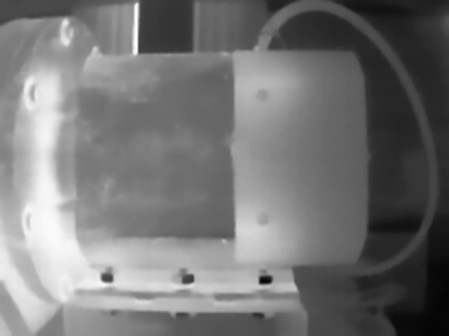}
}
\caption{These figures illustrate the effect of the emissivity of an object's surface on temperature measurements. In the figure, (a)(b) are two bottles of the same shape and different materials, which contain water of the same temperature but present temperature differences in the infrared image; the blower in (c)(d) are composed of different materials. While the internal temperature is homogeneous, the temperature distributions are different in the two regions with different materials in the infrared image.  }
\label{fig:1}
\end{figure}

A critical challenge in infrared thermometry lies in emissivity variability, which introduces measurement inaccuracies when objects with differing emissivities exhibit identical temperatures \cite{cite3}, as shown in Fig.\ref{fig:1}. Traditional approaches rely on manual emissivity input or surface coatings to standardize radiation characteristics \cite{cite4}, but these methods falter with heterogeneous materials. Recent developments propose physics-aware computational corrections, such as Bayesian inference-based temperature calibration \cite{cite33}, which addresses emissivity uncertainties through probabilistic modeling. Concurrently, dynamic emissivity regulation techniques using advanced materials \cite{cite4} offer promising avenues for real-time adaptability.

\begin{figure*}[!t]
\centering
\includegraphics[width=5.6in]{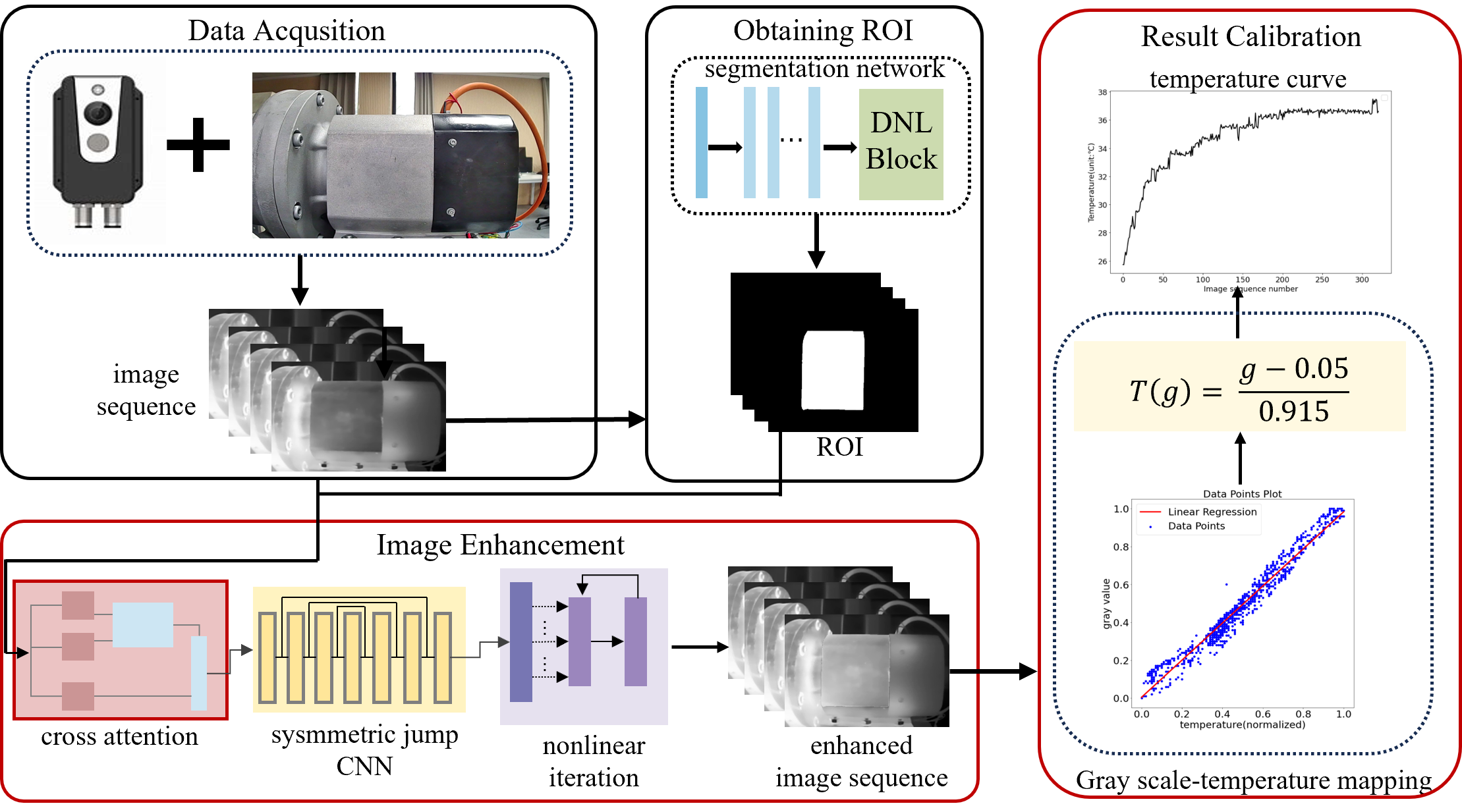}
\caption{
A schematic presentation of the whole processing steps of the proposed method. A sequence of IR images are used as input. They are segmented to obtain RoIs, which are used to focus on two area of the images (composed of Al and Fer). As it is assumed that the inside temperature is the same for the two area, the gray levels of the Al area are modified to have the same statistical properties (means, variance) than the Fer area. Then, a relation between the gray levels and the real temperature is establishes and its parameters are estimated. Having this relation, a temperature value is obtained for each image frame. The evolution of this temperature over frames (time) is plotted. The comparison of this curve with the same before correction, shows the main objective of this work.
}
\label{fig:2}
\end{figure*}
 
Infrared image enhancement has emerged as a complementary strategy to mitigate emissivity artifacts. While classical methods like histogram equalization \cite{cite5} and Retinex algorithms \cite{cite6, cite7} laid the foundation, deep learning now drives significant progress. Techniques such as diffusion model-based enhancement (Diff-Retinex \cite{cite19}) and generative adversarial networks (UW-GAN \cite{cite22}, PUGAN \cite{cite23}) demonstrate remarkable success in restoring thermal and underwater imagery by integrating physical models. For industrial applications, recent work by Liu et al. \cite{cite6} introduced a reflectance-corrected Retinex framework for carbon fiber defect detection, highlighting the synergy between physical priors and data-driven enhancement. These advancements motivate the integration of adaptive enhancement with emissivity calibration for robust thermal diagnostics.

In recent years, research methods have tended to achieve only one of temperature calibration or image enhancement, making it difficult to achieve both at the same time. Methods that can realize both image correction and temperature calibration have not been extensively studied.

This study proposes an thermal infrared image-driven temperature calibration framework for blower systems (Fig.\ref{fig:2}), addressing three key gaps in existing methodologies: (1) emissivity-induced pseudo-color distortions in multi-material components, (2) inadequate adaptive enhancement for localized thermal gradients, and (3) limited contextual fusion between reference and target regions. Building on DNLNet's segmentation capabilities \cite{cite8}, our architecture employs cross-attention mechanisms to fuse regional features, followed by nonlinear iterative enhancement guided by a statistically informed loss function. The innovations are threefold:

\begin{enumerate}
    \item Emissivity normalization: eliminating the differences caused by different surface emissivities, determine the emissivity well \cite{ASTM} in advance, and unify the pixel values of image areas with different emissivities to the same level, make the radiation values reflected by the image more consistent and comparable, and improve the accuracy of the enhanced image.
    \item Local area adaptive enhancement: Based on the probability density difference between the reference area and the target area, as well as feature statistics such as mean, variance, and histogram distribution, a new loss function is designed to perform adaptive enhancement intensity adjustment, so that the details of the target area are better presented without affecting the naturalness of the overall image.
    \item Feature extraction: Using the cross-attention mechanism before image enhancement can effectively fuse the features of the target area and the reference area, thereby better capturing the relationship between the various parts of the image. This preprocessing method can provide richer contextual information for subsequent image enhancement, making the enhancement effect more natural and consistent.
\end{enumerate}

The framework's efficacy is validated through industrial case studies, demonstrating superior accuracy in temperature mapping compared to histogram-based and Retinex-based \cite{cite6} methods. 

\section{Related Works}
\subsection{Image Enhancement}
\subsubsection{Frequency Domain}
Frequency domain-based image enhancement involves transforming an image from the spatial domain to the frequency domain, manipulating its frequency components to enhance specific image features, and then transforming the result back to the spatial domain. This approach leverages spectral information to perform enhancement tasks such as smoothing, sharpening, and denoising. Compared to spatial domain methods, frequency domain techniques are more effective in isolating and enhancing specific image components, especially in applications requiring contrast enhancement and noise suppression.

In particular, wavelet domain-based enhancement techniques have demonstrated notable performance. For instance, one method employs directional bases and frequency-adaptive shrinkage to effectively suppress noise in color images \cite{cite9}, while another approach balances perceptual quality and distortion in super-resolution tasks through wavelet domain style transfer \cite{cite10}. For low-light image enhancement, frequency-based models have been proposed to restore low-frequency object information while enhancing high-frequency details \cite{cite11}. Image quality is further improved through dual-interval contrast enhancement and integrated color correction \cite{cite12}. Specialized applications, such as underwater image enhancement, benefit from a two-step domain adaptation framework \cite{cite13}, and zero-reference diffusion models have been developed for low-light scenes \cite{cite14}. Super-resolution performance has also been improved through cross-domain residual networks \cite{cite15}, while haze removal has been addressed using block-based multi-scale frequency enhancement \cite{cite16}.

Building upon these developments, frequency domain-based infrared image enhancement has further evolved with the introduction of multi-scale decomposition and advanced noise suppression techniques. While traditional wavelet transforms are limited by dynamic range compression issues, recent methods integrate physical models to improve adaptability and perceptual quality. For example, Hao et al. \cite{cite34} proposed a wavelet-based algorithm combined with improved bilateral filtering, which adaptively adjusts decomposition levels to suppress high-frequency noise while preserving edge sharpness in thermal images—achieving a 28\% increase in PSNR over conventional methods. Additionally, the Feature Multi-Scale Enhancement and Adaptive Dynamic Fusion Network (FMADNet) \cite{cite35} employs a Residual Multi-Scale Feature Enhancement (RMFE) module to extract hierarchical features from infrared small targets in complex backgrounds, reaching a 94.5\% recall rate on the NUDT-SIRST dataset. These innovations overcome the limitations of global frequency manipulation by introducing adaptive, locally responsive mechanisms tailored to the structural characteristics of infrared imagery.

Overall, frequency domain-based techniques continue to prove effective in image enhancement, particularly in fields such as medical and infrared imaging. Nevertheless, these methods may introduce artifacts or distortions and often involve significant computational complexity, necessitating careful algorithm design and resource allocation.

\subsubsection{Spatial Domain}
Spatial domain-based image enhancement involves directly manipulating pixel values to adjust brightness, contrast, color, sharpness, and other visual attributes, thereby improving the perceptual quality of the image.

Histogram equalization enhances dynamic range and contrast by redistributing grayscale values uniformly. Feature extraction techniques based on co-occurrence matrices, wavelets, and Gabor filters have been applied to RGB images \cite{cite17}. Retinex-based approaches, such as the super-Laplacian prior \cite{cite18} and the Diff-Retinex model \cite{cite19}, have also been developed to improve visual appearance.

With the advancement of machine learning and deep learning, spatial domain enhancement methods have become increasingly adaptive. RSCNN has been proposed for low-light remote sensing tasks \cite{cite24}, while a modified Unet architecture has demonstrated effectiveness in enhancing low-light images \cite{cite25}. An intelligent bearing diagnosis framework incorporates image enhancement and improved CNN to enhance feature extraction \cite{cite29}.

Infrared imagery, often characterized by low resolution and contrast, has benefited from recent innovations. A cross-modal transformer (CMTR) was introduced to enhance feature extraction across modalities \cite{m-media1}, and MFFENet employed multi-scale fusion for RGB-thermal scene parsing \cite{m-media2}. Local entropy mapping has also been used to improve infrared image quality \cite{cite30}.

Recent advancements further improve spatial methods by integrating models and architectures. Pang et al. \cite{cite36} introduced a Two-Stream Deep Convolutional Neural Network (TS-DCNN) for infrared image enhancement, which combines a detail enhancement sub-network with mixed attention blocks and a content-invariant sub-network based on dilated convolutions to suppress background noise while preserving spatial structure, achieving a 12\% improvement in SSIM over traditional methods.

Compared to frequency domain methods that may introduce artifacts and lose local temperature fidelity, spatial domain techniques better preserve structural and thermal details. Building on Zero-DCE \cite{cite14}, our proposed method incorporates emissivity normalization across heterogeneous materials and introduces an attention mechanism to enhance feature extraction and temperature consistency.


\subsection{Infrared thermal radiation model}
Thermal infrared sensors estimate an object's surface temperature by detecting its emitted thermal infrared radiation. However, the radiated energy is influenced by the object's emissivity: at a constant temperature, lower emissivity results in less emitted energy. Additionally, measured temperature is affected by environmental factors such as ambient temperature, humidity, and atmospheric conditions. To ensure that the enhanced temperature accurately reflects the true surface temperature, this study draws on a Bayesian inference-based calibration method \citep{cite33} to first correct the temperature values recorded by the thermal infrared sensor. Based on the infrared thermal radiation model \citep{cite3}, the actual surface temperature of an object can be computed using Eq.~(\ref{eq:12}):
\begin{equation}
T_r = \left\{ \frac{1}{\epsilon} \left[\frac{1}{\tau_a} \cdot T_m^n - (1 - \epsilon) T_b^n - \left(\frac{1}{\tau_a} - 1\right) T_a^n \right] \right\}^{\frac{1}{n}}
\label{eq:12}
\end{equation}
where $\epsilon$ is the surface emissivity of the object to be measured, $T_m$ is the temperature measured by the thermal infrared imager, $T_r$ is the actual temperature of the object to be measured, $T_b$ is the background temperature, $T_a$ is the atmospheric temperature, $n$ is the constant related to the sensor material, and the value of $n$ for the HgCdTe (8-14 $\mu_m$) sensor is 4.09. $\tau_a$ is the atmospheric transmittance, which can be calculated:
\begin{equation}
\tau_a = e^{-k(h_\omega)d},\quad k(h_{\omega}) = \frac{\omega \times h_{\omega}}{6.76 \times 100\%} \times 0.342
\label{eq:13}
\end{equation}
where $\omega$ is the condensation number of water, the value of $\omega$ at different temperatures is shown in Table \ref{tab:1}, $h_\omega$ is the ambient relative humidity, $k(h_\omega)$ is the atmospheric attenuation coefficient, and $d$ is the distance between the IR sensor and the object \citep{khw}.

\begin{table}[]
\centering
    \caption{Condensation number of water $\omega$} at different temperatures
    \begin{tabular}{cccccc}
    \toprule
     Temperature & 5 & 10 & 15 & 20 & 25\\
     \midrule
     $\omega(mm/km)$ & 6.76 & 9.33 & 11.96 & 17.22 & 22.80\\
     \bottomrule
    \end{tabular}
    \label{tab:1}
\end{table}

\section{Methodology}
This study presents an thermal infrared sequence temperature calibration framework addressing emissivity variations through an optimized enhancement architecture, comprising three principal components:
\begin{enumerate}
    \item \textbf{Image preprocessing}: a segmentation network and mask processing extract the target and reference areas from each image; 
    \item \textbf{Image enhancement}: Mask-guided features undergo cross-attention fusion, followed by multi-scale CNN feature extraction and pixel-wise transformation through a nonlinear iterative module;
    \item \textbf{Temperature calibration}: the enhanced images are converted into a temperature profile using radiation calibration.
\end{enumerate} 

\subsection{Image preprocessing module}
The objective of this subsection is to accurately extract and differentiate material-specific regions in thermal images, providing a reliable foundation for subsequent emissivity normalization and image enhancement. During the material-specific region extraction phase for wind blower components, each sequence image undergoes initial processing through the DNLNet architecture to generate semantic segmentation masks. These label images identify the specific locations of different material regions of the blower. Subsequent morphological operations produce binary masks isolating both the target region (ROI) and reference zones through connected-component analysis.

In this way, we get the image that keeps only the region of interest and other regions become black, the result is shown in Fig.\ref{fig:4}. The material differentiation specifically addresses aluminum and iron within the blower assembly, corresponding to chromatic coding in the segmentation maps (green: aluminum, yellow: iron) based on their characteristic infrared signatures.

\subsection{Image Enhancement Network}

In this part, we adaptively enhance the target region in thermal images by fusing features from both target and reference areas, thereby improving temperature consistency and image quality across materials with different emissivities. The image enhancement network in this paper consists of three parts: cross attention module, CNN and nonlinear iteration. The specific structure is shown in Fig.\ref{fig:12}:
\begin{figure}[!t]
\centering
\includegraphics[width=3.2in]{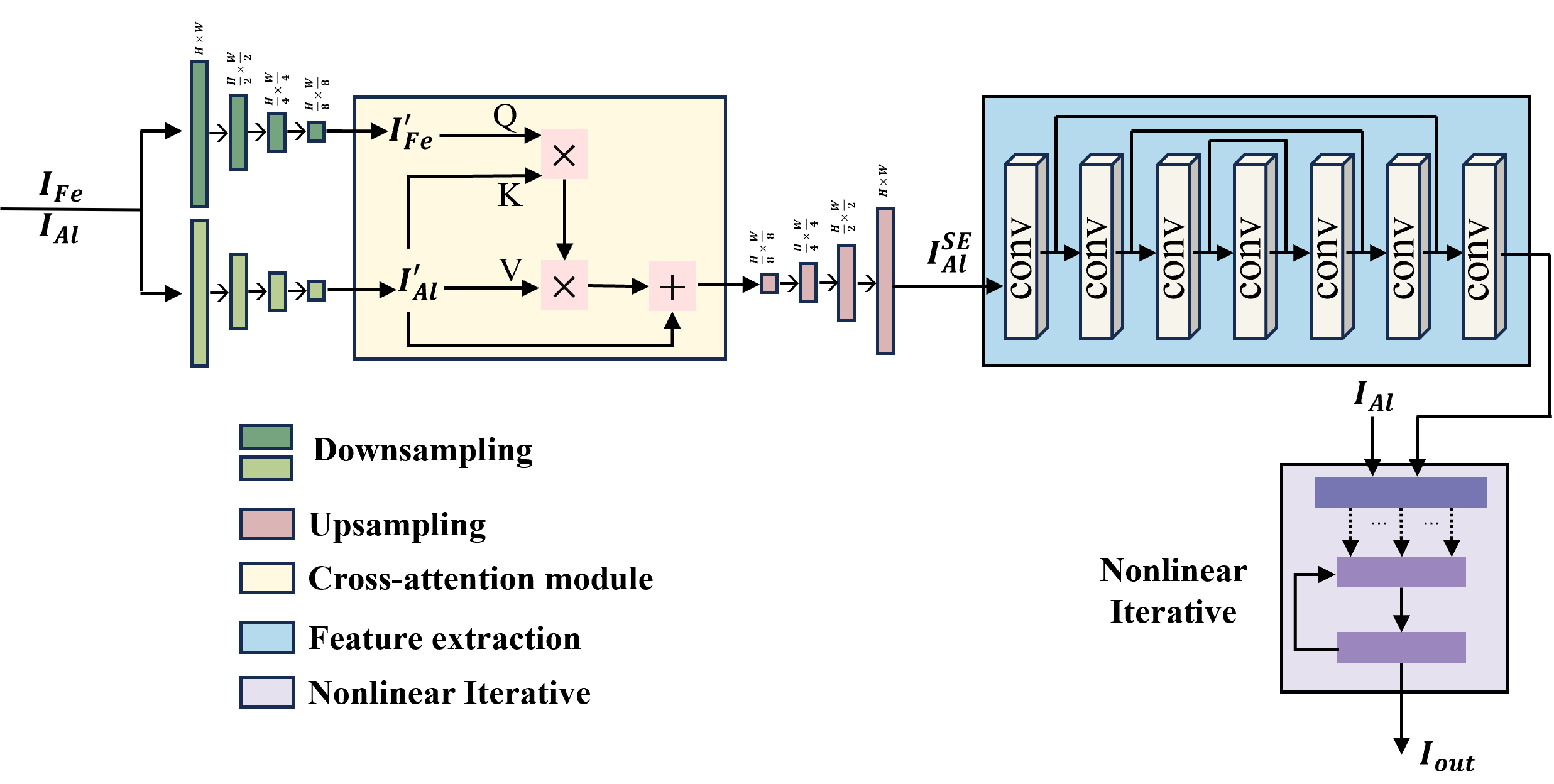}
\caption{Schematic diagram of the image enhancement network structure. This network is an improved Zero-DCE. The input of the network is two images containing only the reference area ($I_{Fe}$) and only the target area ($I_{Al}$). A cross attention module is added before the skip convolutional neural network, and the nonlinear iteration module performs adaptive enhancement based on the grayscale of the original target area.}
\label{fig:12}
\end{figure}

After the mask images of the aluminum area and the iron area are obtained through the segmentation network and mask processing, the two images are converted to the YCbCr color space and then normalized according to the emissivity of the two areas as shown in Eq.(\ref{eq:2}):

\begin{equation}
I_R' = \tanh\left(\frac{I_R}{\varepsilon_R}\right)
\label{eq:2}
\end{equation}
where $R$ refers to aluminum or iron, $I_R$ is the masked image of the $R$ region after conversion to YCbCr color space, $\varepsilon_R$ is the emissivity of the $R$ material, and $I_R'$ is the masked image feature of the processed $R$ region. The emissivities $\varepsilon_R$ are measured in advance for each material \cite{ASTM}.

Following preliminary feature conditioning, hierarchical downsampling operations are executed through four successive stages to progressively reduce spatial resolution while capturing multi-scale hierarchical features. This multi-scale feature extraction enables the network to capture both global context and fine details. The cross-attention mechanism operates with Query (Q) derived from aluminum-associated feature maps, while Key (K) and Value (V) originate from iron representations within the downsampled hierarchy. This architecture enables inter-material correlation modeling through spectral characteristics alignment, mathematically formalized in Eq.(\ref{eq:3}):
\begin{equation}
I_{\text{Al}}^{SE} = \text{softmax}\left(\frac{{QK}^T}{\sqrt{d_k}}\right){V}
\label{eq:3}
\end{equation}
where $\mathbf{Q}$ is the feature map of the aluminum region after four downsampling stages, and $\mathbf{K}$, $\mathbf{V}$ are the feature maps of the iron region after the same process; $\sqrt{d_k}$ is the scaling factor.

The combined two-region feature is upsampled to restore the feature map, \(I_{Al}^{SE}\), and input into a convolutional neural network. The network has seven layers: the first six use 32 3×3 kernels, and the last uses eight 3×3 kernels, all with a stride of 1 and ReLU activation. Skip connections fuse information from different layers to prevent gradient vanishing and feature loss. The 4th and 3rd layer features are concatenated before the 5th layer, and the 5th and 2nd before the 6th. The last layer outputs an 8-channel feature map. 
\begin{equation*}
\begin{aligned}
    & Y_1 = \text{ReLU}(W_1 \cdot X + b_1), \\
    & Y_i = \text{ReLU}(W_i \cdot Y_{i-1} + b_i), \quad i = 2, 3, 4 \\
    & Y_j = \text{ReLU}\left(W_j \cdot \text{Concat}(Y_{8-i}, Y_{i-1}) + b_j\right), \quad j = 5, 6, 7
\end{aligned}
\end{equation*}
where $X$ is the input of the first convolutional layer, i.e., $I_{Al}^{SE}$; $Y_i$ and $Y_j$ are the outputs of the $i$-th and $j$-th convolutional layers, respectively, with $i$ being 2,3,4 and $j$ being 5,6,7; $W$ and $b$ are the weight and bias of each layer, respectively, and $\text{Concat}$ is the tensor concatenation operation.

\begin{figure}[!t]
\centering
\subfigure[iter = 8 with varying $\theta$]{
\includegraphics[width=0.46\linewidth]{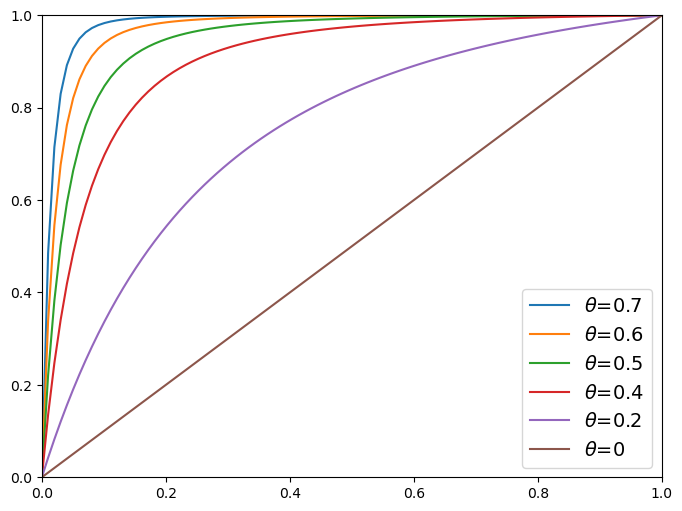}
}
\subfigure[$\theta$ = 0.7 with varying iters]{
\includegraphics[width=0.46\linewidth]{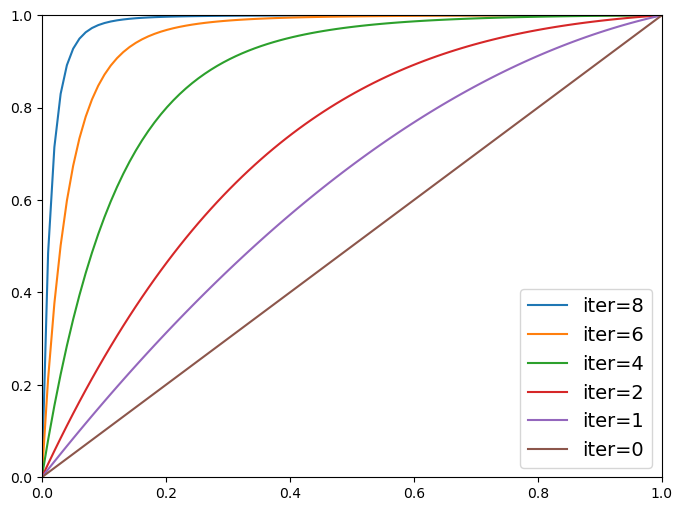}
}
\caption{
This figure demonstrates the effect of different numbers of iterations and different values of $\theta$ on the enhancement results. Here, iter is the number of iterations, and \boldmath$\theta$ is the enhancement strength parameter.
}
\label{fig:5}
\end{figure}

The input image's gray value is adjusted through a nonlinear iterative process to enhance regional images, with the goal of improving local contrast and making the target region's statistical properties more consistent with the reference region. When iteration values approach 0 or 1, the gray value adjustment decreases, while mid-range values increase contrast. Parameters for each iteration are derived from the convolutional network's output, allowing adaptive enhancement based on local features. This ensures proper enhancement of grayscale values across regions. As shown in Fig.\ref{fig:5}, more iterations and larger parameters lead to stronger enhancements. The nonlinear iteration process is described by Eq.(\ref{eq:7}):

\begin{equation}
\begin{aligned}
    & C_0(x) = I_{\text{Al}}', \\
    & C_n(x) = C_{n-1}(x) + \theta_n(x) C_{n-1}(x) \left(1 - C_{n-1}(x)\right), \\
    & \text{where } n = 1, 2, \ldots, 8.
\end{aligned}
\label{eq:7}
\end{equation}
where $C_0 (x)$ is the adjusted aluminum region of Eq.(\ref{eq:2}), $C_n (x)$ denotes the output of the nth nonlinear iteration, and $\theta_n (x) $is the parameter of the nth iteration, where $\theta$ for each iteration is obtained by splitting the output of the convolutional neural network. These operations are performed on the pixel values of the normalized target region after segmentation and emissivity normalization. Here, $\theta_n(x)$ controls the strength of the nonlinear enhancement at each iteration; larger values of $\theta$ result in stronger contrast enhancement in the target region.

\subsubsection{Loss Function}
To better train the image enhancement network, we use the statistical alignment loss $L_{stat}$ and the probability density distance loss $L_{hist}$, with the total function $L_{total}$ as shown in Eq.(\ref{eq:9}):
\begin{equation}
L_{\text{total}} = L_{\text{stat}} + L_{\text{hist}}
\label{eq:9}
\end{equation}

Direct enhancement of image gray values often ignores the physical meaning: temperature-radiance-emissivity, while statistical features such as mean and variance as indirect descriptions of the temperature distribution can avoid over-stylization of the image and enhance the realism of the enhanced image.

The statistical alignment loss $L_\text{stat}$ minimizes discrepancies between the enhanced target region's mean ($\mu_{\text{En}}$) and variance ($\sigma_{\text{En}}$) and the reference region's ($\mu_{\text{Fe}}$, $\sigma_{\text{Fe}}$) to align their statistical distributions, as defined in Eq.(\ref{eq:10}).
\begin{equation}
L_{\text{stat}} = \frac{1}{2} \left[(\mu_{\text{En}} - \mu_{\text{Fe}})^2 + (\sigma_{\text{En}} - \sigma_{\text{Fe}})^2\right]
\label{eq:10}
\end{equation}

Assuming a homogeneous internal temperature and negligible material thickness, surface temperature differences are primarily caused by emissivity variations. To address this, we aim to make the gray-level distributions in target and reference region similar after enhancement. Rather than modifying pixels individually, we adopt a more robust approach by by aligning their statistical distributions: either using the mean and variances alignment by optimizing the criterion (Eq.\ref{eq:10}) or by minimizing the Kullback-Leibler (KL) divergence  \citep{cite31}.

The probability density distance loss $L_{hist}$ minimizes gray-level distribution discrepancies between augmented target and reference regions using KL divergence, which is an asymmetric measure for probability distribution alignment.
\begin{equation}
L_{\text{hist}} = \frac{1}{2} \left[\text{KL}(H_{\text{En}} \parallel H_{\text{Fe}}) + \text{KL}(H_{\text{Fe}} \parallel H_{\text{En}})\right]
\end{equation}

By combining these two loss functions, an enhanced image that is more consistent with the reference region in multiple dimensions can be generated, improving the overall visual quality of the image.
\begin{figure}[!t]
\centering
\subfigure[origin image]{
\includegraphics[width=1in]{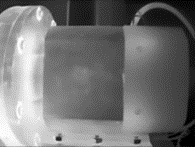}
}\hspace{5mm}
\subfigure[segment labels]{
\includegraphics[width=1in]{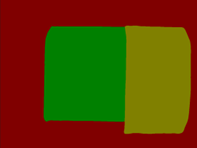}
}\hspace{5mm}
\subfigure[mask of the focused segment]{
\includegraphics[width=1in]{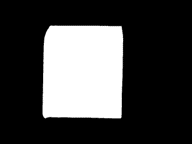}
}\hspace{5mm}
\subfigure[masked area filled with pixel values]{
\includegraphics[width=1in]{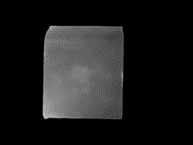}
}
\caption{This figure shows the details of the first step of the processing in Fig.1. Masked image extraction process for different material regions.}
\label{fig:4}
\end{figure}

\subsection{Radiation calibration and temperature conversion module}

Here, we convert the enhanced image’s grayscale values into accurate temperature profiles through radiation calibration, ensuring that the processed images reflect true surface temperatures for quantitative analysis. The enhanced image must visually match the actual temperature distribution, and its temperature values should align with the true temperatures. To verify this, grayscale values are converted to temperature using radiation calibration results.

\begin{figure}[!t]
\centering
\includegraphics[width=2.5in]{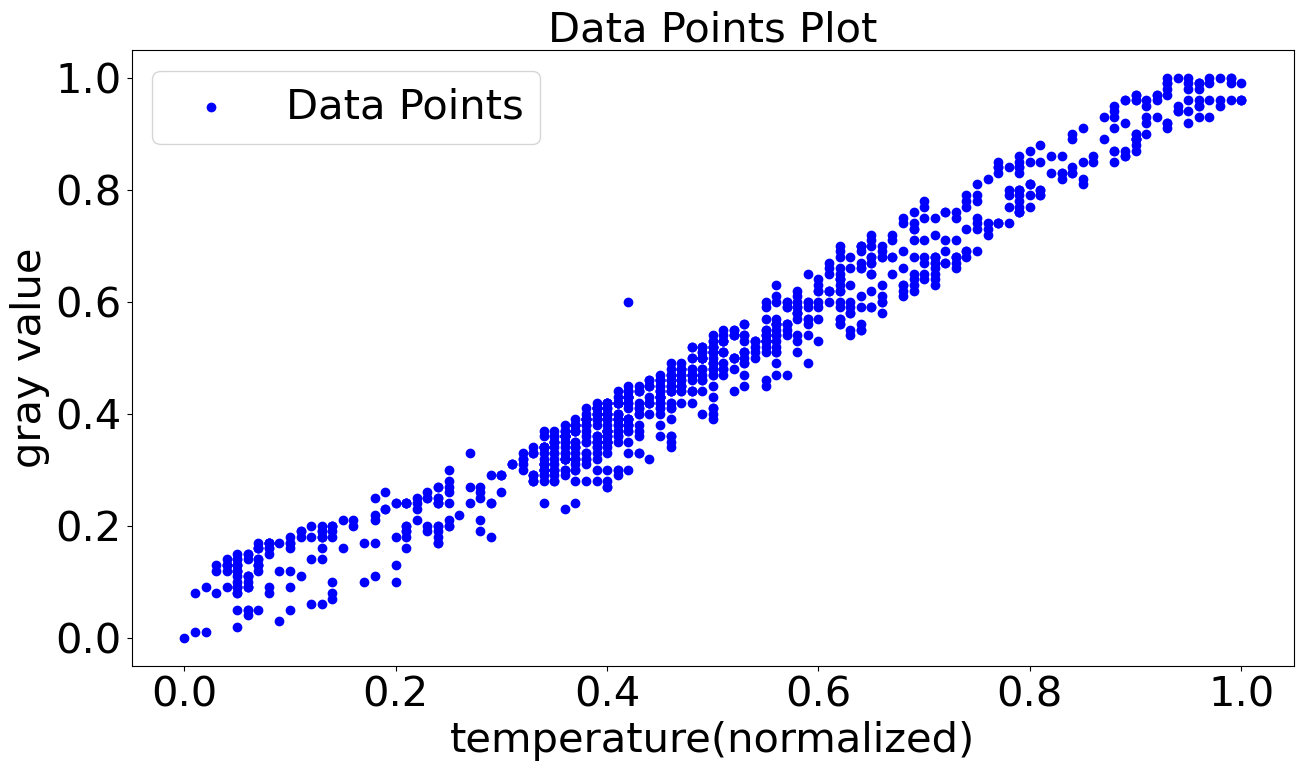}
\caption{This figure shows example of gray-scale-temperature scatter plot.}
\label{fig:3}
\end{figure}

The infrared radiation model is used to calculate the actual GT temperature. Different infrared sensors have varying grayscale-temperature relationships. To determine this relationship for the sensor used in the experiment, the temperature and pixel value files are normalized and plotted as a scatter plot in two-dimensional space (Fig.\ref{fig:3}). By observing the scatter plot, a fitting function is selected and solved using the least squares method.

\begin{figure}[!t]
\centering
\includegraphics[width=3in]{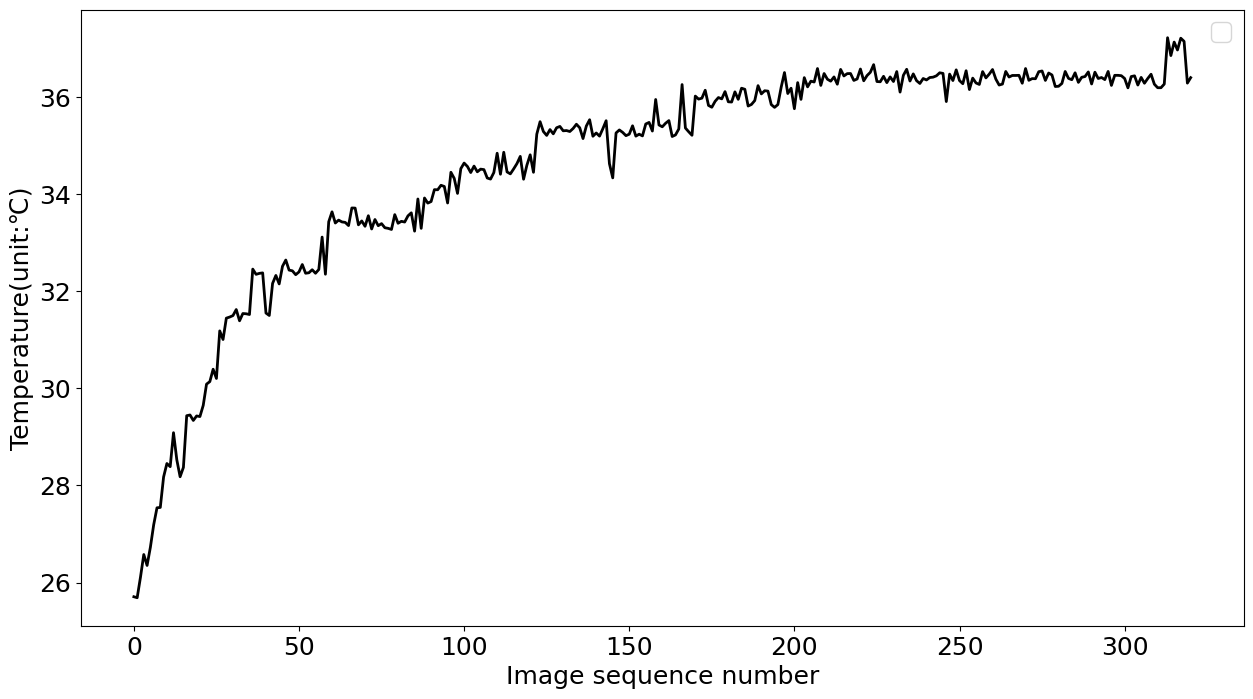}
\caption{Example plot of sequence temperature. Each data point represents the average temperature of the target area in the sequence image. In this experiment the temperature is increasing from 26 at the startin point of recording to about 37 at the end of recording. }
\label{fig:6}
\end{figure}

The temperature of the enhanced area is converted into a normalized temperature value through the result of radiation calibration, and the sequence temperature of the sequence image is obtained. Then the normalized enhanced temperature is restored to the original temperature range through the GT temperature. Fig.\ref{fig:6} shows an example of the sequence temperature diagram.

\section{Experiments and Results}
The thermal infrared sensor used in this experiment is IRay AT20, and the collected data includes thermal infrared image data with a resolution of 1024×768 and the corresponding temperature data of 256×192. In order to make the pixel values of the image data correspond to the temperature data one by one, the bilinear interpolation method with a scaling factor of 0.25 is used for downsampling to obtain an image of 256×192, with each cell corresponding to one pixel.

This paper verifies the effectiveness of the improved network for temperature calibration through two sets of experiments. Each set of experiments starts recording from the start of the blower, and saves and records the image and corresponding temperature data every 5 seconds until the blower temperature stabilizes. The first set of experiments has a shooting distance of 0.25m, an ambient temperature of 23.5$^{\circ}\text{C}$, an ambient humidity of 88.1\%, an atmospheric temperature of 22$^{\circ}\text{C}$, a blower speed of 19000RPM, and a blower running time of 26.75min, with a total of 321 pairs of image temperature data; the second set of experiments has a shooting distance of 0.74m, an ambient temperature of 31.2$^{\circ}\text{C}$, an ambient humidity of 65.5\%, an atmospheric temperature of 33$^{\circ}\text{C}$, a blower speed of 10000RPM, and a blower running time of 21.92min, with a total of 263 pairs of image temperature data.

The minimum amount of data required for network training is only about 300 images, which greatly reduces the complexity of the experiment.

The emissivity measurements of the equipment used in the experiment took contact thermometry, so that other factors such as background had little effect on the accuracy of the results.

\subsection{Radiation Calibration Results}
In an image, a single pixel may be affected by noise, resulting in deviation from the actual value. In order to avoid the influence of noise on radiation calibration, the sliding window size is set to 5×5, the step size is 7, and the image and temperature values are smoothed in the same way. The obtained temperature value and grayscale value are normalized and used as the horizontal and vertical coordinates of the two-dimensional coordinate system, respectively, and the smoothed data points are plotted. The example diagram is shown in Figure \ref{fig:4}. It can be observed from the figure that the normalized grayscale is linearly related to the temperature. In order to further confirm the expression of the objective function, this paper uses the least squares method to solve the specific expressions of different objective functions, and calculates the error between the corresponding value of the objective function ($y_{pred}$) and the true grayscale value ($y$). The error results are shown in Table \ref{tab:objFunc}.

\begin{table*}[ht]
    \centering
    \caption{Mean square error between different forms of objective function($y_{pred}$) and the true gray value ($y$)}
    \resizebox{\linewidth}{!}{

    \begin{tabular}{ccccc}
    \hline
       Objective Function & $y_{pred}=ax+b$ & $y_{pred}=ax^2+bx+c$ & $y_{pred}=ax^3+bx^2+cx+d$ & $y_{pred}=\frac{1}{1+e^{ax+b}}$ \\
        \hline
        MSE & 0.0019 & 0.0016 & 0.0015 & 0.0019 \\
         \hline
    \end{tabular}
    }
    \label{tab:objFunc}
\end{table*}

It can be seen from the Table \ref{tab:objFunc} the different forms of objective functions listed can all solve suitable parameters to make the objective function conform to the relationship between grayscale and temperature, and the mean square error between the objective functions in different forms and the true value is small enough. In order to improve the efficiency of the calculation and simplify the problem, combined with the scatter plot, the objective function is selected as a linear function, and the least squares method is used to solve the parameters that minimize the residual sum of squares:
\begin{equation}
\begin{aligned}
    & G(t) = at + b, \\
    & \text{arg} \min_{(a, b)} \sum_{i=1}^m \left[ g_i^{\text{real}} - (a t_i + b) \right]^2
\end{aligned}
\end{equation}
where $G(t)$ is the objective function, a,b are the parameters to be solved, $t_i$ is the $i$th temperature value, $g_i^\text{real}$ is the actual gray value corresponding to the $i$th temperature value, and m is the total number of temperature values.

The specific expression of the objective function is obtained by calibration:
\begin{equation}
    G(t) = 0.915t+0.05
\end{equation}

\begin{figure}[!t]
\centering
\includegraphics[width=0.9\linewidth]{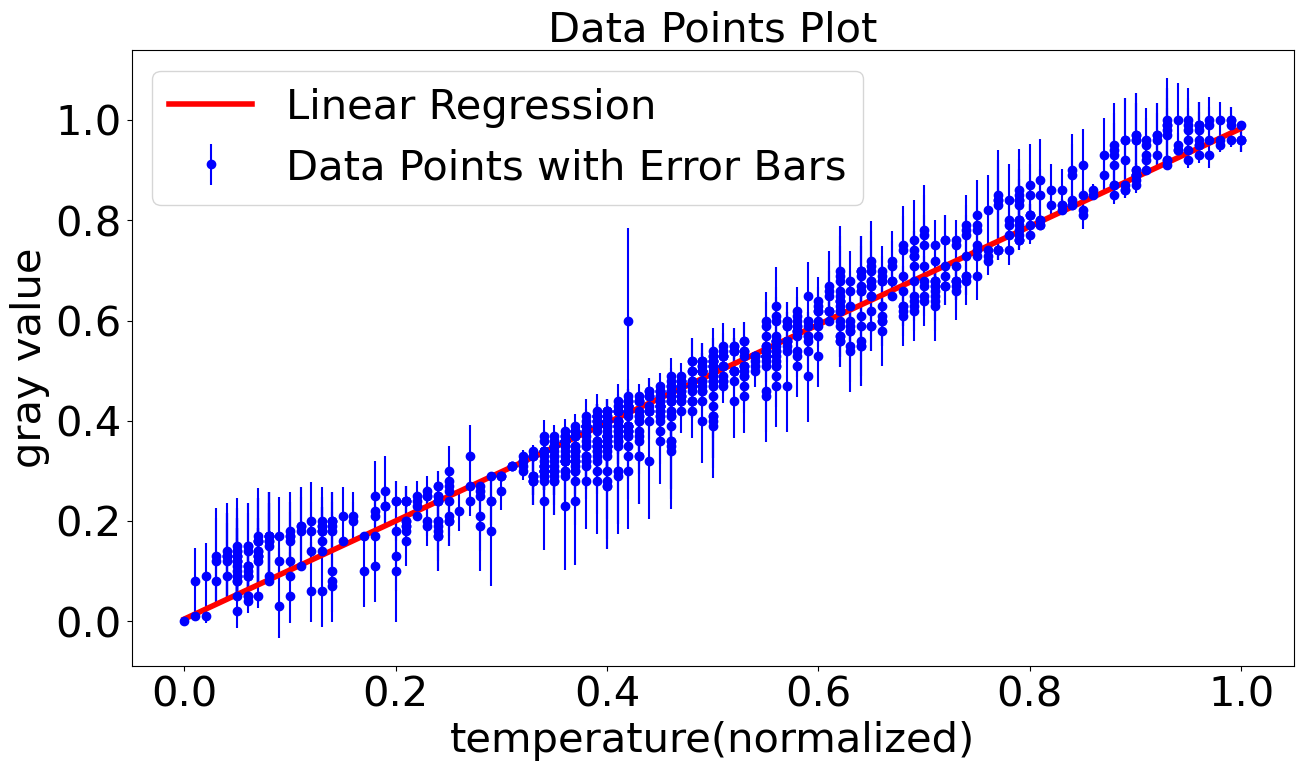}
\caption{Image of radiation calibration results.The red line is the regression line of the fitted grayscale-temperature correspondence.}
\label{fig:7}
\end{figure}

The regression fitting graph is drawn in the scatter plot, as shown by the red line in Fig.\ref{fig:7}(b). The mean square error (MSE) between the calibrated result and the actual data point pixel value is 0.00234.  

\subsection{GT Temperature}
\begin{figure}[!t]
\centering
\includegraphics[width=\linewidth]{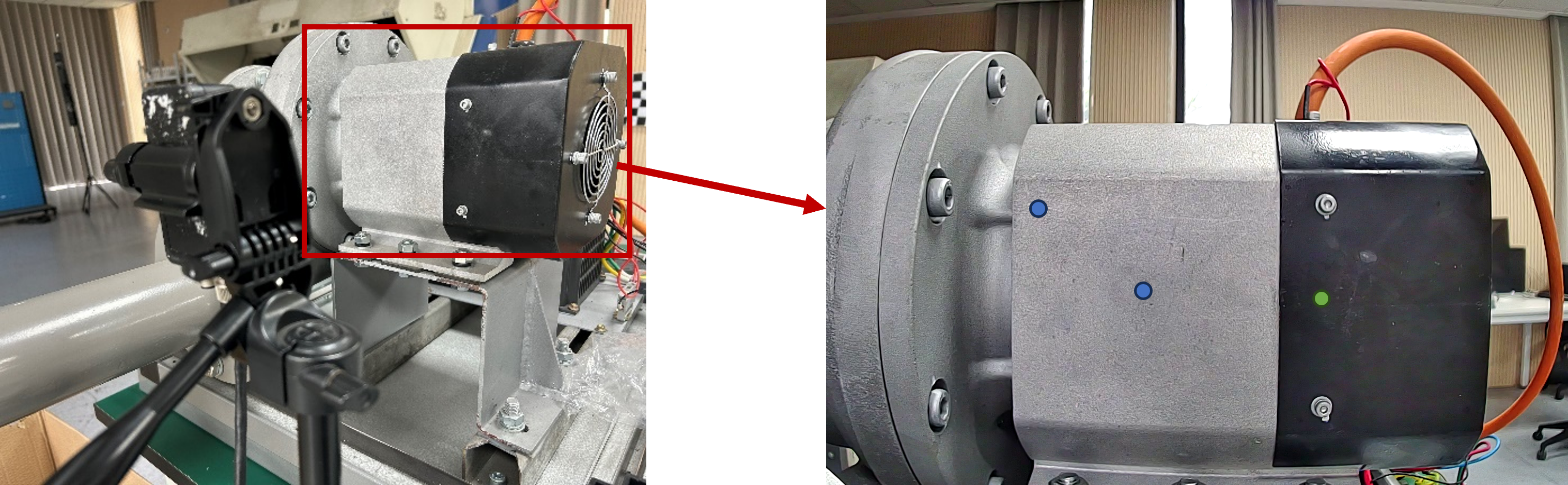}
\caption{Photography of the blower and calibration points.}
\label{fig:8}
\end{figure}

Three calibration points (two aluminum, one iron) were designated on the blower surface, as shown in Fig.\ref{fig:8}.

\begin{figure}[!t]
\centering
\includegraphics[width=0.95\linewidth]{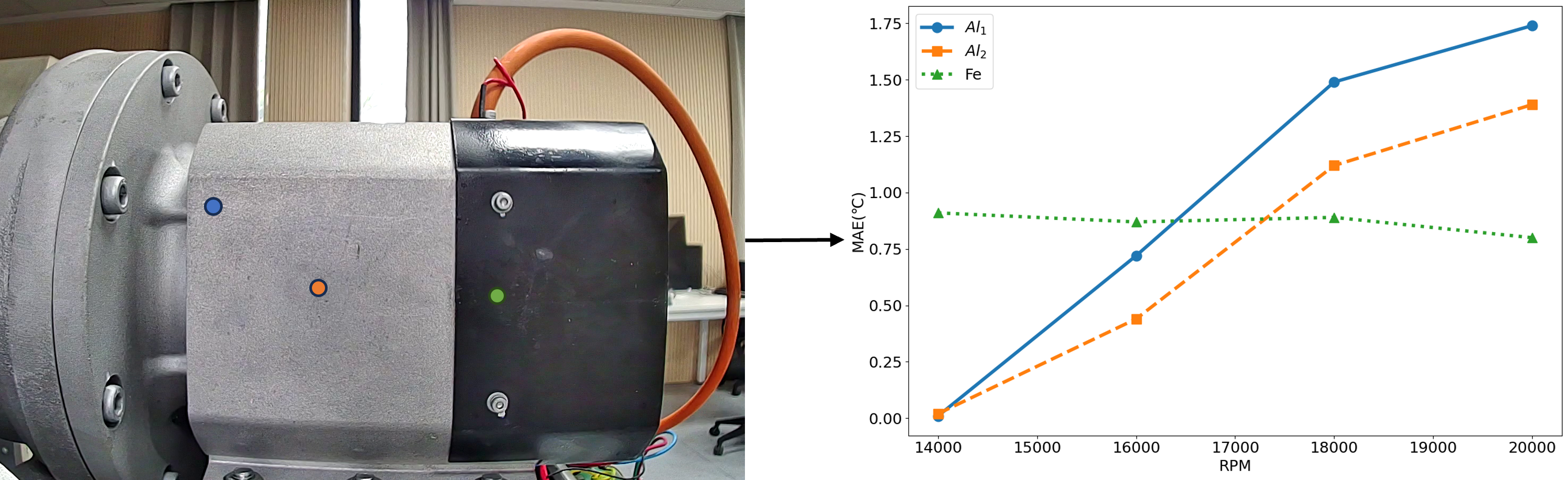}
\caption{MAE between the calibrated temperature and the actual surface temperature of the calibration point at different rotational speeds.}
\label{fig:9}
\end{figure}

Set the blower speeds to 14,000–20,000 RPM (2,000 increments), turn on the blower and wait for stabilization. Record the experimental conditions such as ambient temperature, humidity, background temperature and atmospheric temperature. When the internal temperature of the blower stabilized, the actual surface temperature at the calibration point was measured and recorded. The temperature data saved by the sensor was calibrated by Eq.(\ref{eq:12}) and compared with the actual temperature and the results are shown in Fig.\ref{fig:9}. Error magnitudes increased with rotational speed yet remained within industrial acceptability thresholds ($\le$ 20,000 RPM), validating the calibrated temperatures as reliable GT references.

\subsection{Sequence image enhancement results}
\begin{figure}
    \centering
    \subfigure[]{\includegraphics[width=0.32\linewidth]{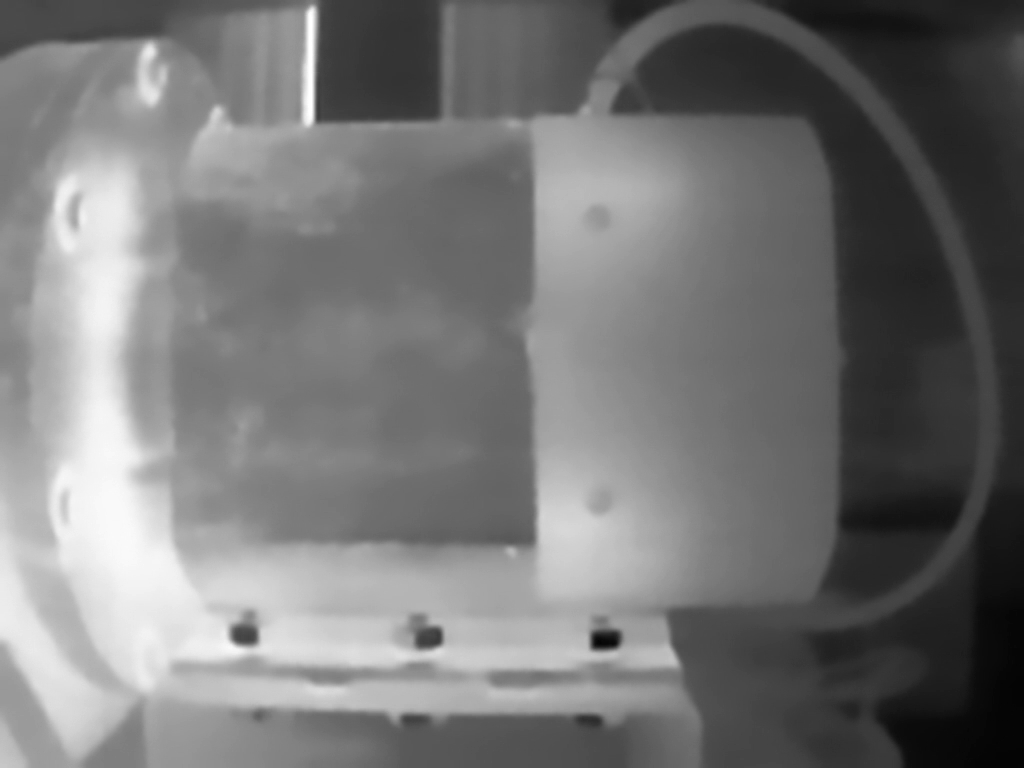}}
    \subfigure[]{\includegraphics[width=0.32\linewidth]{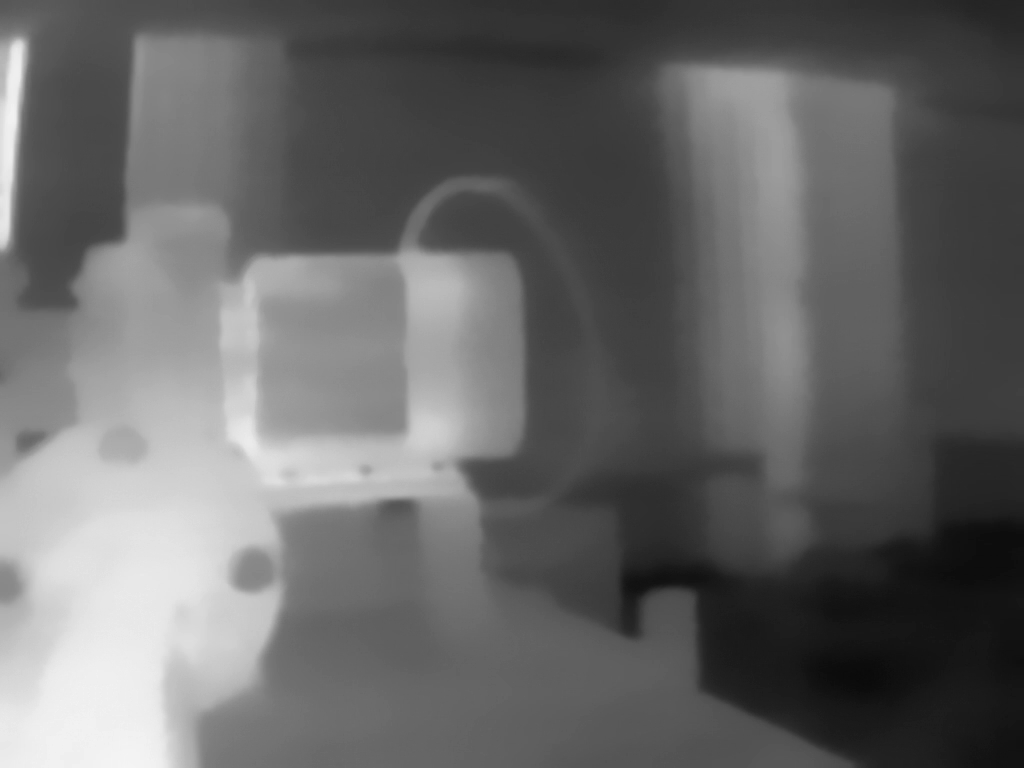}}
    \subfigure[]{\includegraphics[width=0.32\linewidth]{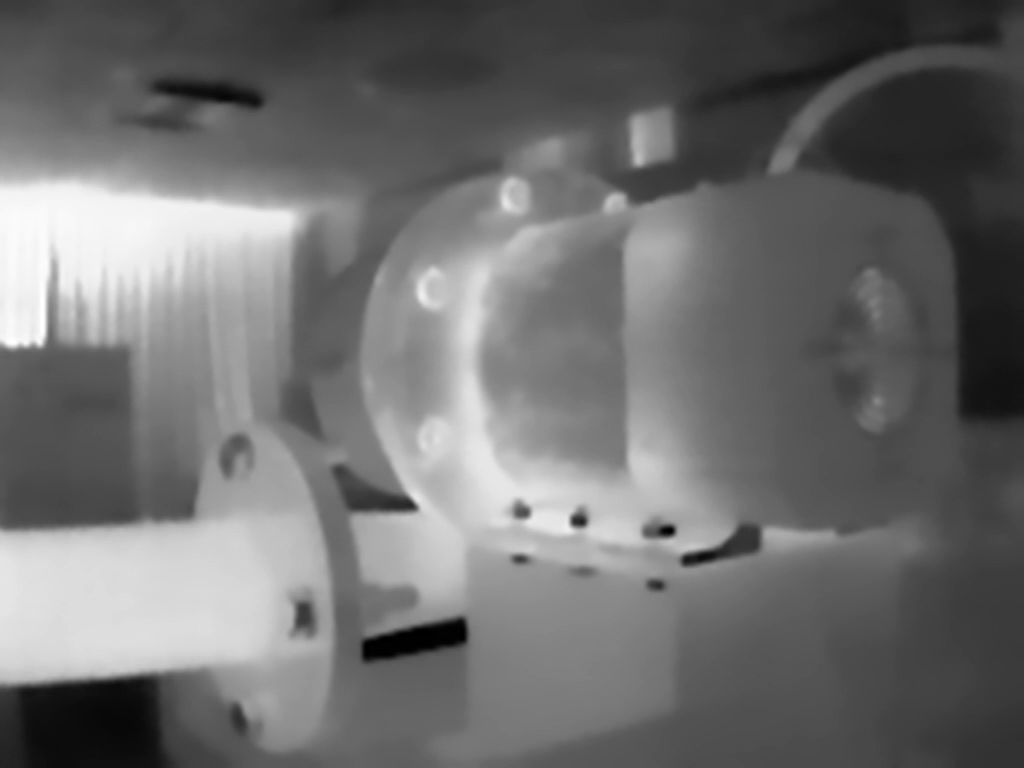}}
    \caption{Some examples of experiments at different distances and angles. (a) and (b) at two different distances, and (c) at a different angle.}
    \label{fig:9.5}
\end{figure}

The experiments verified the plausibility of the enhancement results from different environmental conditions, such as multiple distances and multiple angles, as shown in Fig.\ref{fig:9.5}. Two sets of these experiments are shown in this section.

\begin{figure*}
\vspace{-10pt}
\centering
\resizebox{\textwidth}{!}{
    \begin{tabular}{p{0.7in}ccc|p{1in}}
          Originals&  \includegraphics[width = 1in]{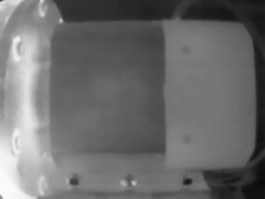} &  \includegraphics[width = 1in]{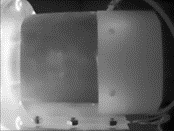} &  \includegraphics[width = 1in]{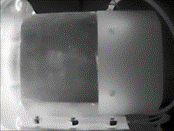} &  \multirow{2}{*}{\includegraphics[width=1in]{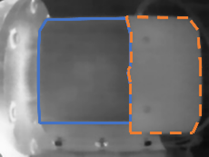}} \\
         
        Enhanced &\includegraphics[width = 1in]{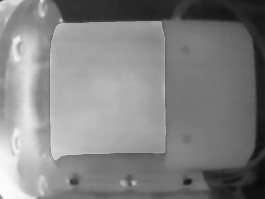} &  \includegraphics[width = 1in]{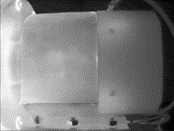} &  \includegraphics[width = 1in]{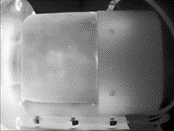} &   \\
      \cline{1-5}
      \\
      Originals&  \includegraphics[width = 1in]{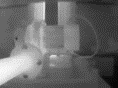} &  \includegraphics[width = 1in]{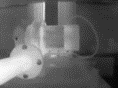} &  \includegraphics[width = 1in]{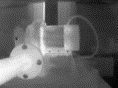} &  \includegraphics[width=1in]{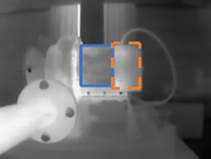}\\
         
        Enhanced &\includegraphics[width = 1in]{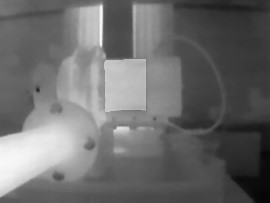} &  \includegraphics[width = 1in]{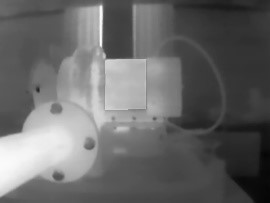} &  \includegraphics[width = 1in]{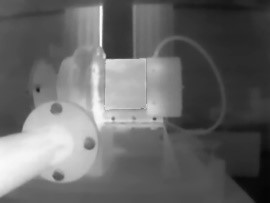} &  \includegraphics[width=1in]{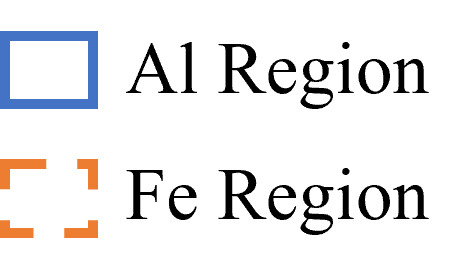} \\
        \hline
        Blower's running time & 1min & 7min & 26min & Description of materials
    \end{tabular}
    }
\vspace{-7pt}
\caption{Images after running the blower for different times with corresponding enhanced images. }
\label{fig:10}
\end{figure*}

Fig.\ref{fig:10} shows the results before and after image enhancement of two sets of images at different times during the operation of the blower, the first set of images of the blower taken at a distance of 0.3 m, and the second set of images of the blower taken at a distance of 0.75 m. There is a significant improvement in the contrast of the enhanced images visually.
In order to evaluate the enhancement results more intuitively, the average values of the metrics were calculated for the sequential images of the blower's operation process, and the pairs of metrics are shown in Table \ref{tab:2}.
\begin{table}[ht]
    \centering
    \caption{Comparison of image metrics before and after enhancement}
    \begin{tabular}{cccc}
    \toprule
     & SSIM$\uparrow$ & CEI$\uparrow$ & Entropy$\uparrow$ \\
    \midrule
    Experiment 1 & 0.9420 & 1.344 & 7.4460 \\
    Experiment 2 & 0.9915 & 1.095 & 7.2379 \\
     \bottomrule
    \end{tabular}
    \label{tab:2}
\end{table}

SSIM is used to measure the similarity between the two images, and the SSIM before and after enhancement are 0.9420 and 0.9915, respectively, indicating that the enhanced images have high similarity in brightness, contrast and structure. CEI measures the contrast of the enhanced image with respect to the original image, and the CEI is 1.344 and 1.095, indicating that the contrast of the enhanced image is improved. Entropy is 7.4460 and 7.2379, the range is usually 0-8, higher values indicate that the enhanced image contains more details and information.

\begin{figure*}[!t]
\centering
\subfigure[Experiment 1]{
\includegraphics[width=0.46\linewidth]{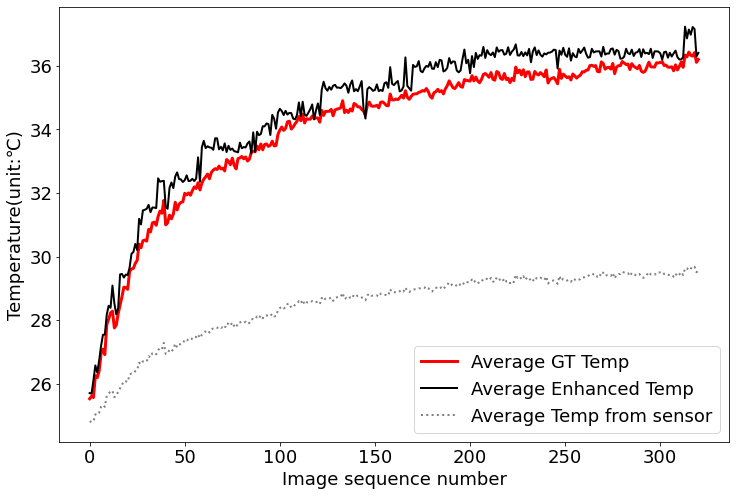}
}
\subfigure[Experiment 2]{
\includegraphics[width=0.46\linewidth]{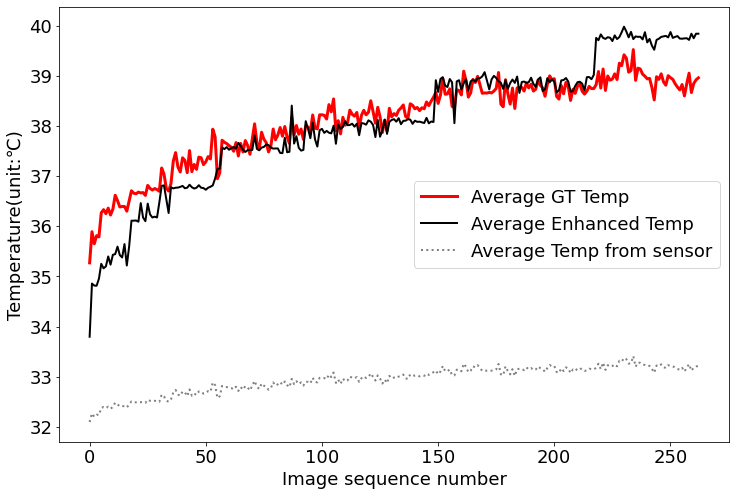}
}

\caption{Temperature profile of enhanced temperature, GT temperature and sensor reading temperature. The results of the temperature profiles for the two sets of experiments correspond to the two sets of experiments in Fig.\ref{fig:10}.}
\label{fig:11}
\end{figure*}

The goal of image enhancement is not only to restore the visual temperature distribution but also to ensure that the enhanced temperature matches the true value. To compare the enhanced image temperature with the actual temperature, the normalized enhanced temperature is calibrated using the GT temperature and reduced to the original value. The 95\% maximum GT temperature is set as the upper limit and the minimum value is set as the lower limit. A 16×16 neighborhood is randomly selected to calculate the average temperature distribution. Figure \ref{fig:11} shows the comparison of the average temperature: red (enhanced), black (GT), and gray (original).

As can be seen from the figure, in the first set of experiments, the temperature after image enhancement was very close to the GT temperature in the average results for randomly selected regions. In the second set of experiments, the enhanced images in the early stage of blower operation have a large error of close to 1$^{\circ}\text{C}$ from the GT temperature, but in the late stage of blower operation, the enhanced temperature is similar to the GT temperature. In order to compare the difference between the enhanced temperature and the GT temperature more intuitively, we use the Euclidean distance to calculate the distance between the enhanced temperature and the original temperature and GT temperature profiles, and the Euclidean distance formula is shown in Eq.(\ref{eq:17}).
\begin{equation}
    \text{Dis} = \sqrt{\sum_{i=1}^{N}  (T_i^{En} - T_i^{GT})^2}
    \label{eq:17}
\end{equation}
where $T^{En}$ denotes the post-enhancement temperature, $T^{GT}$ denotes the GT temperature, $N$ is the total number of sequence images, and $i$ denotes the index of the sequence images.

The results of the Euclidean distance calculations are shown in Table \ref{tab:3}. The error of the post-enhancement and GT temperatures are shown in Table \ref{tab:4}.

\begin{table}[ht]
    \centering
    \caption{The distance between the original temperature profile, the enhanced temperature profile and the GT temperature profile}
    \begin{tabular}{ccc}
    \toprule
     & Experiment 1 & Experiment 2 \\
     \midrule
     $Dis(Orig,GT)$  & 102.5069 & 84.1086 \\
     $Dis(En,GT)$ & 11.2188 & 8.2627\\
     \bottomrule
     
    \end{tabular}
    \label{tab:3}
\end{table}

\begin{table}[ht]
    \centering
    \caption{Error between enhanced and GT temperatures}
    \begin{tabular}{ccc}
    \hline
      & Experiment 1 & Experiment 2 \\
     \hline
     ERR($^\circ\text{C}$) & $0.57\pm0.25$ & $-0.02\pm0.51$ \\
     \hline
    \end{tabular}
    \label{tab:4}
\end{table}

Combining quantitative and qualitative analyses, it can be seen that the error between the enhanced sequence temperature and the GT temperature is very small, and thus the present enhancement algorithm achieves both visual and physical corrections.

\section{Conclusions}
This paper proposes an algorithm for temperature calibration of different emissivities regions of thermal infrared sequence images using an improved Zero-DCE-based image enhancement network. The algorithm consists of three parts: preprocessing of thermal infrared images of the target area, image enhancement, and temperature conversion. Different from traditional methods, this paper adopts a virtual substitution approach, image enhancement, and innovatively achieves the calibration of distorted images and temperature. In particular, in the image enhancement module, the emissivity of different materials is processed, and an attention module is introduced. Experimental results show that this algorithm can solve the problem of temperature distortion caused by emissivity differences. The image enhancement algorithm maintains its original temperature distribution while enhancing the image, and does not lose its own temperature distribution information; The results of expanding a single image into a series of images and the comparison of the enhanced image converted to temperature with the GT temperature both prove the accuracy of the enhancement algorithm.

Since the image enhancement method is used for temperature calibration for the first time, the experimental conditions are limited. At present, it is limited to calibrating two materials with different emissivity.  Future efforts will focus on extending this approach to materials with multiple emissivity values to further enhance its applicability. Future work will explore multi-material calibration scenarios ($\ge$ 3 emissivity values) under dynamic environmental conditions, and investigate physics-informed neural architectures to improve cross-domain generalization capabilities for industrial thermal monitoring systems. This method provides a more convenient way to monitor the condition of equipment in industry and provides new ideas for future research. 

This method has broad potential for industrial applications, such as automated substation inspection, wind turbine blade and cable monitoring, and general equipment health management.
By enabling accurate temperature calibration and image enhancement across different materials, our approach improves the efficiency and reliability of fault detection and predictive maintenance in real-world scenarios.
The ability to incorporate prior knowledge of material emissivities further enhances its adaptability and practical value for intelligent equipment monitoring.




\printcredits

\bibliographystyle{cas-num}

\bibliography{references}



\end{document}